1# A Novel Approach in Power System State Estimation Based on a System-State Controller

Ramin Faraji
Inamori school of Engineering
Alfred University
rf8@alfred.edu

*Abstract*—This paper summarizes an optimal state estimation that previously has been used in power systems and discusses the robustness of the by using a system-state controller. Static state estimation (SSE) traditionally tries to linearize power systems' measurement functions in a simple Gauss-Newton method optimization problem to obtain the best estimation of our system states. However, SSE would not be able to respond to system dynamics very well since it ignores to consider system nonlinearities. A state-space controller and a new solution method is deployed to overcome that difficulties. A set of simulation results is provided to confirm the suitability of the solution.

## I. INTRODUCTION

ACCESS to highly reliable electricity is one of the most crowd-pleasing matters in the future digital world. Along the same line, certain methods can enhance the robustness of power system reliability to a considerable degree. For example, the author in [1] has reinforced the self-healing characteristic of the system by a novel method of optimally allocating control and protective devices in the system, which drastically improved the reliability of the system. Furthermore, state estimation can lead to improving the reliability of the power systems by another method. In other words, control of the power system parameters needs to estimate the system states as accurately as possible [2]. Brand-new challenges for both planning and operating power systems that electrical engineers are facing include:

1- Conventional static state estimation (SSE) cannot absorb the fast and stochastic changes [3] in the transmission and distribution power system [4].
2- Current measurement tools are not fast enough to capture and detect the aforementioned stochastic changes [3].

To solve the first issue, we should use either a dynamic state estimation (DSE) or a robust to SSE [5]. However, in the second issue it is stated that in practice, it is not possible to have measurements that are fast enough to deploy DSE [6]. Practically, a robust SSE could be proposed to obtain an accurate enough estimation of our system states.

Conventionally in SSE, we might hire a simple WLS optimization to minimize the difference between the actual measurements and the system measurement functions value (mainly power equations using voltage angle and magnitude [7]) by estimating the closest set of state variable vectors through linearization progress. However, the main problems with SSE are that it may not consider state transitions [8], there are neither memories of previous states nor prediction of future states [9], and it may only consider the spatial aspect of the system [10].

To come up with an appropriate solution, a system state-space controller using Bellman equation and an approximated solution for it has been deployed to overcome nonlinear source difficulties.

This paper tries to robust state estimation by using a state-space model controller for a power system [13]. An actual 14-bus IEEE power system would be treated with this controller and the results of the simulation that proves that this robustness may help the control of power systems in terms of confronting fast and stochastic change [3].

Furthermore, the rest of the paper is dedicated to as the following: section II contributes to problem statement. The control part is presented in section III. The solution approach is discussed in section IV. A case study simulation is presented in section V, and lastly, conclusions are drawn in section VI.

## II. PROBLEM FORMULATION

There are many studies on power system state estimation and they consist of either static or dynamic approaches [5].

In order to have a robust system state estimation against fast and stochastic changes in power-flow estimation, the dynamic sort of estimation has recently received much attention [6]. In order to accomplish this in practice, we do need to have both fast and digital measurements along with control and processing units with high speed processing and communication rates in parallel [11], [12]. However, many kinds of these dynamic estimations are not applicable in real-world power systems due to a lack of required measurement and processing units.

Here is the formulation for a simple estimation+4 without any constraints in the power system:

$$z = \begin{bmatrix} z_1 \\ z_2 \\ \vdots \\ z_m \end{bmatrix} = \begin{bmatrix} h_1(x_1, x_2, \dots, x_n) \\ h_2(x_1, x_2, \dots, x_n) \\ \vdots \\ h_m(x_1, x_2, \dots, x_n) \end{bmatrix} + \begin{bmatrix} e_1 \\ e_2 \\ \vdots \\ e_m \end{bmatrix} = h(x) + e \quad (1)$$

where:
$h^T = [h_1(x), h_2(x), \dots, h_{m(x)}]$
$h_i(x)$ is the nonlinear function relating measurement $i$ to the state vector $x$.
$x^T = [x_1, x_2, \dots, x_n]$ is the system state vector.
$e^T = [e_1, e_2, \dots, e_m]$ is the vector of measurement errors.

The problem of optimal static estimation is formulated in this section in order to have a minimum amount of residual error. The objective function of this problem is shown in (2)

The WLS estimator will minimize the following objective function:

$$J(x) = \sum_{i=1}^{m} \frac{(z_i - h_i(x))^2}{R_{ii}}$$
$$= [z - h(x)]^T R^{-1}[z - h(x)] \quad (2)$$
$$R = diag\{\sigma_1^2, \sigma_2^2, \ldots, \sigma_m^2\}$$

The standard deviation $\sigma_i$ of each measurement $i$.

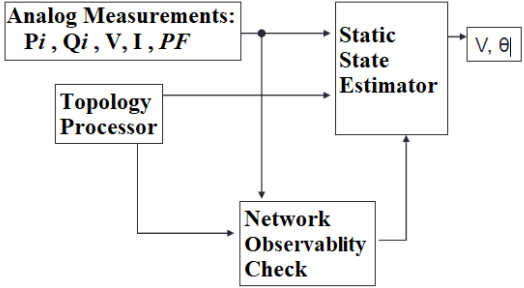

Figure 1. Conceptual way of conventional SSE

If we consider the X vector as voltage magnitudes and angles and any h(x) as a power-flow equation, the optimization can be solved through (3):

$$x^{k+1} = x^k - [G(x^k)^{-1} \cdot g(x^k)] \quad (3)$$

where
  $k$ is the iteration index,
  $x^k$ is the solution vector at iteration $k$,
  $G(x^k) = \frac{\partial g(x^k)}{\partial x} = H^T(x^k). R^{-1}. H(x^k)$
  $g(x^k) = -H^T(x^k). R^{-1}. (z - h(x^k))$.
  Where $H(x) = [\frac{\partial h(x)}{\partial x}]$
  $x^T = [\theta_2, \theta_3, \ldots, V_1, V_2, \ldots]$

To update the X vector in each iteration, we must use (4):

$$\Delta x^{k+1} = H^T(x^k)R^{-1}[z - h(x^k)].[G(x^k)]^{-1} \quad (4)$$

where
$\Delta x^{k+1} = x^{k+1} - x^k$

For more details on this, reference [14].

*State-Space Model of the System*

For any system, state-space model can be formulated as:

$$\dot{x}(t) = ax(t) + bu(t), \quad (5)$$

Where

$$\dot{x}(t) = \begin{bmatrix} \theta_i(t) \\ V_i(t) \end{bmatrix}, \quad u(t) \in \{0,1\},$$

Consequently, the difference between estimated valve and its actual value can be stated as:

$$q(x_k) = (x_k - r)^T Q(x_k - r), \quad (6)$$

where $Q$ is an $n \times n$ positive semidefinite matrix.

For the scalar output:

$$y_k = hx_k, \quad k = 0,1,2,\ldots. \quad (7)$$

### III. CONTROLLER DESIGN

Fig. 2 shows how a control loop can be designed for a system. Then, a control method solution for the problem has been presented In Section III.

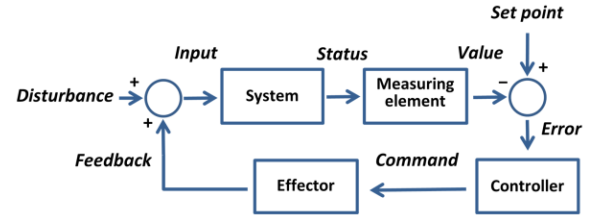

Figure 2. a closed-loop control system

#### A. Bellman Equation

With defining a Bellman equation for the estimation problem of a bounded cost per stage [13]. To rewrite The Bellman equation:

$$V(x,z) = \min_{u \in \{0,1\}} \{c(x,z,u) + \alpha V(Ax + bu, u)\}, \quad (8)$$

where $V(x,z)$ is the value function, a new notation is defined as

$$V_i(x) = V(x,i), \quad i \in \{0,1\} \quad (9)$$

where $V_i(x)$ is the value function while it considers another value of $i$.

To minimize the value of cost function we can substitute $u$ with couple of terms. Thus:

$$V(x,a) = q(x) + \min\{\beta z + \alpha V_0(Ax), \beta(1-z) + \alpha V_1(Ax + b)\}. \quad (10)$$

To shorten the equation, we can define:

$$f(x) = V_1(Ax + b) - V_{0(Ax)}. \tag{11}$$

*B. Find a Solution*

To solve the equations, we can use [12] method, so:

$$V_0(x) = q(x) + \min\{\alpha V_0(Ax), \beta + \alpha V_1(Ax + b)\},$$

$$V_1(x) = q(x) + \min\{\beta + \alpha V_0(Ax), \alpha V_1(Ax + b)\} \tag{12}$$

Making the equation to a symmetrical form, we can rewrite this as:

$$V_0(x) = q(x) + \alpha V_0(Ax) + \min\{0, \beta + \alpha f(x)\},$$

$$V_0(x) = q(x) + \beta + \alpha V_1(Ax + b) + \min\{-(\beta + \alpha f(x), 0\}. \tag{13}$$

To eliminate the minimizer in the upper part we can rewrite the equation as:

$$V_0(x) = q(x) - \tfrac{1}{2}|\beta + \alpha f(x)| + \tfrac{1}{2}(\beta + \alpha V_0(Ax) + \alpha V_1(Ax + b)),$$

$$V_1(x) = q(x) - \tfrac{1}{2}|\beta - \alpha f(x)| + \tfrac{1}{2}(\beta + \alpha V_0(Ax) + \alpha V_1(Ax + b)). \tag{14}$$

Reconsidering the quadratic approach, we can rewrite the equation as:

$$V(x) = (x - \theta)^T P(x - \theta) + v, \tag{15}$$

where $\theta$ is a constant $n \times 1$ vector, $v$ is a small positive value, and $P$ is a constant $n \times n$ positive definite matrix. With replacing the following terms as in last equation, we can consider:

$$P = Q + \alpha A^T P A,$$

$$\theta = P^{-1}(I - \alpha A^T)^{-1}\left(Q_r - \tfrac{1}{2}\alpha A^T P b\right),$$

$$v = \tfrac{1}{1-\alpha}\left(r^T Q r + \tfrac{1}{2}\beta + (\alpha - 1)\theta^T P \theta + \tfrac{1}{2}\alpha b^T P b - \alpha b^T P \theta\right). \tag{16}$$

Verifying the feasibility of mentioned solution needs a calculation of $f(x)$:

$$f(x) = x^T \delta + \zeta, \tag{17}$$

For a $2 \times 1$ constant vector $\delta$ and a constant scalar $\zeta$:

$$\delta = -2A^T P \theta, \quad \zeta = \theta^T P \theta - 2b^T P \theta.$$

so:

$$f(x) = \sum_{k=1}^{n}(\delta_k x_k - m_k). \tag{18}$$

The main purpose of this paper is to optimize the new residual function for minimizing total errors in the power system. In this work, to analyze the voltage magnitude and angle of each bus, there is a need to run the program iteratively which is very popular for distribution and power transmission systems. As formulated in section III, the optimal control in this paper is a non-linear problem which has been solved using Bellman equation solution method. In addition, solutions via particle swarm optimization (PSO) methods and the heuristic [15],[16] and genetic algorithms are also examined [17],[18].

## IV. CASE STUDY SIMULATION

*A. Case Study*

In order to understand on the state estimation with a state-space model controller, we implement and run power flow from an IEEE 14 bus test power system. Moreover, a set of one snapshot memory is considered in the external iteration to consider the system dynamics. In this paper, power lines' data and buses' data are presented in tables I and II, respectively. In fact, a set of power injections along with 3 different time snapshots of our power system are shown in Fig. 3 and 4 to illustrate our robustness in power system state estimation.

Table I: Parameters of Lines Data

| From Bus | To Bus | R pu | X pu | B/2 pu |
|---|---|---|---|---|
| 1 | 2 | 0.01938 | 0.05917 | 0.0264 |
| 1 | 5 | 0.05403 | 0.22304 | 0.0246 |
| 2 | 3 | 0.04699 | 0.19797 | 0.0219 |
| 2 | 4 | 0.05811 | 0.17632 | 0.0170 |
| 2 | 5 | 0.05695 | 0.17388 | 0.0173 |
| 3 | 4 | 0.06701 | 0.17103 | 0.0064 |
| 4 | 5 | 0.01335 | 0.04211 | 0.0 |
| 4 | 7 | 0.0 | 0.20912 | 0.0 |
| 4 | 9 | 0.0 | 0.55618 | 0.0 |
| 5 | 6 | 0.0 | 0.25202 | 0.0 |
| 6 | 11 | 0.09498 | 0.19890 | 0.0 |
| 6 | 12 | 0.12291 | 0.25581 | 0.0 |
| 6 | 13 | 0.06615 | 0.13027 | 0.0 |
| 7 | 8 | 0.0 | 0.17615 | 0.0 |
| 7 | 9 | 0.0 | 0.11001 | 0.0 |
| 9 | 10 | 0.03181 | 0.08450 | 0.0 |
| 9 | 14 | 0.12711 | 0.27038 | 0.0 |
| 10 | 11 | 0.08205 | 0.19207 | 0.0 |
| 12 | 13 | 0.22092 | 0.19988 | 0.0 |
| 13 | 14 | 0.17093 | 0.34802 | 0.0 |

Table II: Parameters of Buses Data

| Bus | Vsp | PGi | QGi | PLi | QLi |
|---|---|---|---|---|---|
| 1 | 1.060 | 0 | 0 | 0 | 0 |
| 2 | 1.045 | 40 | 42.4 | 21.7 | 12.7 |
| 3 | 1.010 | 0 | 23.4 | 94.2 | 19.0 |
| 4 | 1.0 | 0 | 0 | 47.8 | -3.9 |
| 5 | 1.0 | 0 | 0 | 7.6 | 1.6 |
| 6 | 1.070 | 0 | 12.2 | 11.2 | 7.5 |
| 7 | 1.0 | 0 | 0 | 0.0 | 0.0 |
| 8 | 1.090 | 0 | 17.4 | 0.0 | 0.0 |
| 9 | 1.0 | 0 | 0 | 29.5 | 16.6 |
| 10 | 1.0 | 0 | 0 | 9.0 | 5.8 |
| 11 | 1.0 | 0 | 0 | 3.5 | 1.8 |
| 12 | 1.0 | 0 | 0 | 6.1 | 1.6 |
| 13 | 1.0 | 0 | 0 | 13.5 | 5.8 |
| 14 | 1.0 | 0 | 0 | 14.9 | 5.0 |

## B. Simulation Results

Our problem is simulated via MATLAB program and the result of this simulation is shown in Table III.

Table III: Parameters of the System State Estimation in Times t and t-1 Snapshot

| Bus No. | t Vpu | t Angle (deg) | t-1 Vpu | t-1 Angle (deg) |
|---|---|---|---|---|
| 1 | 1.0068 | 0.0000 | 1.0182 | 0.0000 |
| 2 | 0.9891 | -5.5321 | 1.0012 | -5.5008 |
| 3 | 0.9601 | -14.3214 | 0.9628 | -14..1358 |
| 4 | 0.9611 | -11.3700 | 0.9689 | -11.3602 |
| 5 | 0.9715 | -9.8333 | 0.9725 | -9.7121 |
| 6 | 1.0285 | -16.1791 | 1.0299 | -16.0042 |
| 7 | 0.9989 | -14.7211 | 1.0032 | -14.6808 |
| 8 | 1.0387 | -14.7681 | 1.0401 | -14.6798 |
| 9 | 0.9863 | -16.4225 | 0.9874 | -16.4340 |
| 10 | 0.9788 | -16.5476 | 0.9869 | -16.6683 |
| 11 | 0.9992 | -16.3397 | 1.0044 | -16.4617 |
| 12 | 1.0029 | -17.1203 | 1.0121 | -16.9405 |
| 13 | 0.9979 | -17.2583 | 1.0052 | -16.9783 |
| 14 | 0.9729 | -17.9967 | 0.9757 | -17.8121. |

## V. CONCLUSIONS

In this paper, the robustness of state estimation is analyzed by using a state-space controller for power system. Bellman equation and a solution are hired to model proposed method in the objective function of minimizing the difference between measurement data and the function amount. Some linearization, along with nonlinear functions, are deployed to solve the estimation problem robustness. The results of this robust estimation are shown in comparing table that shows the ability to obtain accurate results. Although SSE might have some deficiencies in the process and control of the power system, it is still a good method for practical purposes. Definitely, robustness in this method of power system estimation can be one of the best solutions to obtain meticulous estimation.

## REFERENCES

[1] Heydari, S., Mohammadi-Hosseininejad, S. M., Mirsaeedi, H., Fereidunian, A., & Lesani, H. (2018). Simultaneous placement of control and protective devices in the presence of emergency demand response programs in smart grid. *International Transactions on Electrical Energy Systems*, *28*(5), e2537.

[2] Narimani, M. R., Asghari, B., & Sharma, R. (2017, November). Energy storage control methods for demand charge reduction and PV utilization improvement. In *2017 IEEE PES Asia-Pacific Power and Energy Engineering Conference (APPEEC)* (pp. 1-5). IEEE.

[3] Parsa Sirat, A. (2018). Loss Minimization through the Allocation of DGs Considering the Stochastic Nature of Units. *MPRA Paper 87636, 2018, University Library of Munich, Germany.*

[4] Saadatmand, S., Sanjarinia, M. S., Shamsi, P., Ferdowsi, M., & Wunsch, D. C. (2019). Neural Network Predictive Controller for Grid-Connected Virtual Synchronous Generator *in North American power symposium (NAPS) 2019. IEEE, 2019, pp. 1–6.*

[5] Aminifar, F., Shahidehpour, M., Fotuhi-Firuzabad, M., & Kamalinia, S. (2013). Power system dynamic state estimation with synchronized phasor measurements. *IEEE Transactions on Instrumentation and Measurement*, *63*(2), 352-363.

[6] Gomez-Exposito, A., Abur, A., Rousseaux, P., de la Villa Jaen, A., & Gomez-Quiles, C. (2011). On the use of PMUs in power system state estimation. *Proceedings of the 17th PSCC*.

[7] Zimmerman, R. D., Murillo-Sánchez, C. E., & Thomas, R. J. (2010). MATPOWER: Steady-state operations, planning, and analysis tools for power systems research and education. *IEEE Transactions on power systems*, *26*(1), 12-19.

[8] Krause, O., & Lehnhoff, S. (2012, September). Generalized static-state estimation. In *2012 22nd Australasian Universities Power Engineering Conference (AUPEC)* (pp. 1-6). IEEE.

[9] Naeimi, H., & Keppel, D. (2018). *U.S. Patent Application No. 10/061,376.*

[10] Niknam, T., Narimani, M. R., Farjah, E., & Bahmani-Firouzi, B. (2012). A new evolutionary optimization algorithm for optimal power flow in a power system involving unified power flow controller. *Energy Education Science and Technology Part A: Energy Science and Research*, *29*(2), 901-912.

[11] Narimani, M. R., Molzahn, D. K., Nagarajan, H., & Crow, M. L. (2018, November). Comparison of Various Trilinear Monomial Envelopes for Convex Relaxations of Optimal Power Flow Problems. In *2018 IEEE Global Conference on Signal and Information Processing (GlobalSIP)* (pp. 865-869). IEEE.

[12] Shahidehpour, M., & Wang, Y. (2004). *Communication and control in electric power systems: applications of parallel and distributed processing*. John Wiley & Sons.

[13] Parsa Sirat, A., & Zendehdel, N. (2020). A New Approach in Optimal Control of Step-Down Converters Based on a Switch-State Controller, In *2020 IEEE Texas Power and Energy Conference (TPEC)*.

[14] Abur, A., & Exposito, A. G. (2004). *Power system state estimation: theory and implementation*. CRC press.

[15] Saadatmand, S., Sanjarinia, M. S., Shamsi, P., Ferdowsi, M., & Wunsch, D. C. (2019). Heuristic Dynamic Programming for Adaptive Virtual Synchronous Generators. *arXiv preprint arXiv:1908.05744*.

[16] Parsa Sirat, A., Mehdipourpicha, H., Zendehdel, N. & Mozafari, H. (2019). Sizing and Allocation of Distributed Energy Resources for Loss Reduction using Heuristic Algorithms. In *2020 IEEE Power and Energy Conference at Illinois (PECI).*

[17] Lawanson, T., Karandeh, R., Cecchi, V., Wartell, Z., & Cho, I. (2018, April). Improving Power Distribution System Situational Awareness Using Visual Analytics. In *SoutheastCon 2018* (pp. 1-6). IEEE.

[18] R. Karandeh, W. Prendergast, V. Cecchi, "Optimal scheduling of battery energy storage systems for solar power smoothing", *South-eastCon 2019*, pp. 1-6, April 2019.